\documentclass[pre,floatfix,twocolumn, 10pt]{revtex4} 
\usepackage{graphicx}
\usepackage{amssymb}
\usepackage{natbib}
\usepackage{dcolumn}
\usepackage{bm}
\usepackage{color}
\usepackage[colorlinks=true, linkcolor=blue, citecolor=blue]{hyperref}
\usepackage{subfigure}
\usepackage{graphicx,psfrag,xspace}
\usepackage{color}
\usepackage{amssymb,amsfonts,amsmath}
\usepackage{setspace}

\begin{document}

\author{E. A. Jagla} 
\affiliation{Centro At\'omico Bariloche, Instituto Balseiro, 
Comisi\'on Nacional de Energ\'ia At\'omica, CNEA, CONICET, UNCUYO,\\
Av.~E.~Bustillo 9500 (R8402AGP) San Carlos de Bariloche, R\'io Negro, Argentina}

\title{Quasi-static deformation of yield stress materials: homogeneous or localized?
}
\begin{abstract} 
We analyze a mesoscopic model of a shear stress material with a three dimensional slab geometry, under an external quasistatic deformation of a simple shear type. 
Relaxation is introduced in the model as a mechanism by which an unperturbed system achieves progressively mechanically more stable configurations.
Although in all cases deformation occurs via localized plastic events (avalanches) we find qualitatively different behavior depending on the degree of relaxation in the model. For no or low relaxation yielding is homogeneous in the sample, and even the largest avalanches become negligible in size compared with the system size (measured as the thickness of the slab $L_z$) when this is increased. On the contrary, for high relaxation the deformation localizes in an almost two dimensional region where all avalanches occur. Scaling analysis of the numerical results indicates that in this case the linear size of the largest avalanches is comparable with $L_z$ even when this becomes very large. We correlate the two scenarios with a qualitative difference in the flow curve of the system in the two cases, which is monotonous in the first case, or of the velocity weakening type in the second case.
 
\end{abstract}

\maketitle

\section{Introduction}

\begin{figure}
\includegraphics[width=7cm,clip=true]{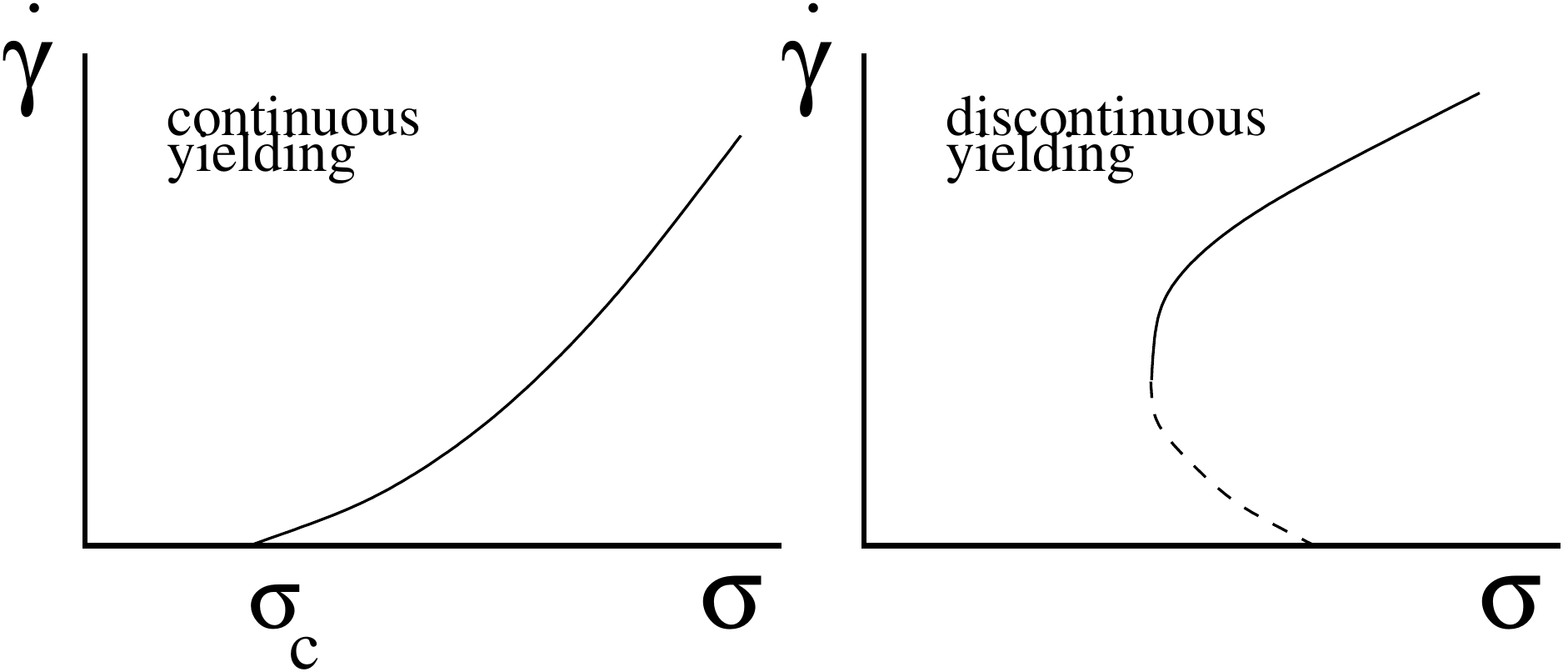}
\caption{Qualitative flow curve of a shear stress material with continuous (a) or discontinuous (b) yielding behavior. The dotted part of the flow curve in (b) is unstable and not experimentally observable, as it is typically replaced by some sort of generic Maxwell construction.}
\label{esquema}
\end{figure}

Yield stress materials \cite{coussot,bonn,nicolas} are substances that share features of both solids and liquids. Under a well defined applied shear stress, known as the critical stress $\sigma_c$, they deform elastically and reversibly. However, if this critical stress is exceeded the material responds plastically, attaining a finite rate of deformation (or strain rate) noted as $\dot \gamma$. 
The dependence of $\dot\gamma$ on $\sigma$ defines the flow curve of the material. 
In a large class of yield stress materials the behavior of $\dot\gamma(\sigma$) is continuous at $\sigma_c$: $\dot\gamma$ increases smoothly from zero when $\sigma$ is increased beyond $\sigma_c$, typically as a power law of the form $\dot\gamma\sim (\sigma-\sigma_c)^\beta$\cite{fisher,kardar}. 
In other cases, there may be a discontinuous transition at $\sigma_c$, typically with some hysteresis\cite{picard,olmsted,divoux,jagla_2007}.
The discontinuous case is also referred to as ``reentrant", as it is typically obtained when the flow curve displays a negative slope region  (Fig. \ref{esquema}) qualitatively similar to the one occurring in a liquid-gas transition in the van der Waals approximation.
To discuss in a unified framework the different phenomenology associated to these two possible scenarios is one of the aims of this work.

Deformation of yield stress materials proceeds via discrete plastic rearrangements generically called ``avalanches" that reduce the shear stress locally, while increasing it in other parts of the sample via anisotropic elastic interactions. In a stress controlled experiment, if $\sigma<\sigma_c$ the material reaches--after a transient--a static equilibrium state in which deformation rate is zero. On the other hand, if
$\sigma>\sigma_c$ there are at any moment plastically active places in the sample that 
produce a finite average value of the deformation rate $\dot\gamma$.

Within the protocol of applying a fixed value of $\sigma$
the identification of isolated avalanches is not possible. However, individual avalanches can be identified and characterized by considering a strain driven experiment, in which the average strain in the sample $\gamma$ is controlled. The value of $\gamma$ is assumed to increase infinitely slowly with time (thus $\dot\gamma\to 0$). In this
scenario, the stress in the sample increases continuously everywhere until there is a place where a mechanical instability occurs (typically a saddle-node bifurcation), and an avalanche is triggered. This avalanche produces a reduction of the stress in the system that can be taken as a measure of the size of the avalanche. 
The collection of points affected by the avalanche can also be determined, defining the avalanche spatial extent.
Upon the further increase of $\gamma$, a sequence of avalanches will be observed and their statistical properties (such as size and duration distributions) can be studied. 

The discrete nature of the deformation of yield stress materials has received a lot of attention in the last years, and in particular the properties of the avalanches responsible for the overall behavior have been studied in detail \cite{lin,karmakar,dahmen,tyukodi}. 
Yet, although the deformation of all yield stress materials can be described in terms of individual avalanches, it is crucial to understand the different effect that these avalanches may have in the behavior of macroscopic samples. There are cases in which the effect of individual avalanches becomes less and less detectable when system size increases. 
This corresponds to a case in which the behavior of a macroscopic sample can be characterized as 
``ductile". In other cases, the effect of individual avalanches continues to be detectable even if the sample size is increased more and more. We will refer to the behavior in such cases as ``fragile"\cite{fragile}.

One of the main points we want to convey in this work is that the distinction between ductile and fragile behavior is intimately related to the underlying flow curve of the material being of the continuous or reentrant type. 
Another important difference between the two cases will emerge from the analysis: while ductile behavior is associated with avalanches that occur all across the sample (therefore producing on average a spatially uniform strain rate), the fragile case produces avalanches that tend to localize spatially in a quasi-two dimensional region where the strain rate remains finite, while it is essentially zero in the rest of the sample. This behavior can be described as the sample developing a ``fault" in it, where all deformation occurs. In fact, a second idea that we want to convey is that 
materials with fragile behavior deform through the production of ``earthquakes" that localize in a sort of ``seismic fault". Beyond this localization effect, an important characteristic that will emerge is that individual earthquakes may have a macroscopically detectable effect, contrary to the case of avalanches in systems without relaxation. This will make connection with the deeply established idea in the geophysical community that in order to sustain earthquakes, a material must have a friction law that is velocity weakening, i.e., a region in which the stress in the system decreases as a function of the strain rate.\cite{scholz} This is precisely the situation with materials having a reentrant flow curve. 

When materials with reentrant yielding are driven at a finite value of $\dot\gamma$ deformation localizes spatially in the form of a shear band\cite{sb1,sb2,sb3,sb4,sb5,sb6}. In fact, a situation reminiscent of coexistence in first order phase transition occurs and there is a spatial separation between a non-flowing region, and a flowing region in the form of a shear band in the system.
The properties of shear bands have received a lot of attention from the materials science community, as they are intimately related for instance to the failure mechanisms of materials like metallic glasses\cite{mg1,mg2,mg3,mg4}. The width of a shear band decreases as the global strain rate decreases, as a kind of Maxwell construction argument easily shows. In the limit of very low deformation rate ($\dot\gamma\to 0$) the thinning of shear bands leads to the formation of a ``fault" where earthquake-like avalanches occur. 
Yet, numerical models devised to study the statistical properties of earthquakes do not consider typically a bulk three dimensional sample. Instead, it is usually assumed from the beginning that the deformation is localized in a two dimensional fault, and therefore only this fault is modeled in detail, perhaps with the surrounding three dimensional medium considered at a mean field level.\cite{bk1,bk2,ofc}
One standard approach is to model the two-dimensional fault as
a collection of blocks that are joined by springs. The velocity weakening feature is incorporated either as a direct external ingredient (as in the Burridge Knopoff model\cite{bk1}, in the form of the local sliding law of each block), or it emerges as a consequence of a more fundamental relaxation mechanism, such as the visco-elasticity model discussed in \cite{landes}.
In either case, the three-dimensionality of the system is incorporated only through the stiffness of a driving spring acting on the effective two-dimensional fault.
In addition, most simulations of two dimensional faults consider the elastic interaction within the fault to occur among ``nearest neighbors" only. Actually, in a real situation this interaction is mediated by the bulk three dimensional medium, and it is long range in an infinite sample. 

In the present work we intend to provide a comparative view of the cases of yield stress materials with continuous or reentrant behavior, and the relation to the earthquake phenomenology. 
We use a three dimensional model, therefore not assuming any simplifying two dimensional situation. We take a system of size $L\times L\times L_z$ (where $L$ is supposed to be ``very large", see below), with periodic boundary conditions, and apply a simple shear deformation of the form ${\bf u}= \gamma z \hat x$, with $\gamma$ defining the strain.
The variation of $L_z$ allows us to perform scaling analysis of the size of the avalanches observed. Note that $L_z$ would eventually relate to the value of the driving spring that simplifying two dimensional model for earthquakes use, and also controls the detailed form of the elastic interaction between different points of the fault.
For each value of $L_z$, the value of $L$ is chosen to be sufficiently large in such a way that the properties of the avalanches are not affected by a further increase of $L$. It is the value of $L_z$ that controls the ``size effects" in the system.

The model (details in the next Section) is driven through the quasistatic increase of the strain $\gamma$. This produces a
a sequence of avalanches that can be characterized in detail. 
In addition, relaxation is introduced through a parameter $R$, and the phenomenology observed depends crucially on the ratio $R/\dot\gamma$ \cite{footnote1}.

Here is a brief summary of the main results that are obtained.
When $R/\dot\gamma\to 0$ we reproduce well known results corresponding to a system with a continuous yielding transition.
In particular, we obtain a distribution of avalanche size of the form $P(S)\simeq S^{-\tau}\exp{(-S/S_{max})}$ with a value of $\tau\simeq 1.5$, and a cut off avalanche size  $S_{max}$ that scales with $L_z$ as
$S_{max}\simeq {L_z}^{1.1}$. The actual observation of the avalanches reveals that although they display the spatial correlations expected from the form of the elastic interaction kernel, they appear all across the sample. Therefore in the long run the deformation of the system is uniform in this case. 

For finite $R/\dot\gamma$, we observe the following trends:

-Avalanches become localized along the $z$ axis, defining a quasi-two-dimensional region that can be called a ``seismic fault". %This localization persists in time,  but we have observed that it may eventually change to a new different position, either abruptly, or by a slow diffusive-like displacement (???). 

-The distribution $P(S)$ becomes broader, with $S_{max}$ increasing with  $R/\dot\gamma$ for a fixed $L_z$. 
The dependence of $S_{max}$ with $L_z$ displays a power law increase with an exponent that becomes larger as $R/\dot\gamma$ increases. For large $R/\dot\gamma$, there are avalanches whose linear size ($\sim S^{1/2}$) becomes larger than $L_z$. This means that avalanches will be macroscopically observable even in the large system size limit ($L_z\to \infty$).

-The stress distribution across the system is qualitatively different for small or large $R/\dot \gamma$. At small values of this parameter, stress fluctuations decrease rapidly in the sample as system thickness increases. Yet for large $R/\dot\gamma$ stress maintains long wavelength spatial fluctuations that decay more slowly with $L_z$.

-An examination of the flow curve of the system reveals a monotonous behavior of $\dot\gamma$ vs $\sigma$ when $R=0$, but clear signs of velocity weakening when $R/\dot\gamma\ne0$. This velocity wakening behavior is seen to be less detectable as the system size (namely $L_z$) increases.

These results make clear the differences between two different yielding behavior of materials, one that can be qualified as smooth, or ductile, and a second one that is fragile, and akin to a system that localizes deformation and can produce earthquakes at the largest scales.

\section{The model}

Our approach is based on the modeling presented in \cite{jagla_2007}.
The applied deformation on the system corresponds to an affine  displacement field $\bf u$ of the form 
${\bf u}\sim (\gamma z,0,0)$, which gives  the ${xz}$ component of the strain as the only one that is non zero. The average strain is therefore $\gamma$, and is the quantity that is externally controlled.  
At the mesoscopic level, the system responds with a local strain $e_{xy}({\bf r})$ that for simplicity will be simply referred to as $e({\bf r})$, and is the main variable in the problem. Note that $\overline {e({\bf r})}=\gamma$, where the overline indicates a spatial average.

The temporal evolution equation for $e({\bf r})$ is assumed to be of the overdamped form, therefore giving $\dot e({\bf r})$
as proportional to a generalized force acting on $e({\bf r})$.
This force has two contributions: a local part $f_{loc}$ and an elastic interaction part $f_{int}$.
The elastic interaction term describes the effect that a change in $e({\bf r})$ has on a different point ${\bf r}'$. It can be expressed as an integral over the whole system  of the form
\begin{equation}
f_{int}({\bf r})=\int d{\bf r'}G({\bf r},{\bf r'})e({\bf r'})
\end{equation}
The elastic kernel $G({\bf r},{\bf r'})$ depends on the symmetry of the applied deformation, and it also includes the effect of other elastic deformations with different symmetry that are note taken into account explicitly in the formalism. In the case in which the deformation is a pure shear, $G$ takes the form that is known as the Eshelby interaction, which has a quadrupolar symmetry. However, for the present simple shear deformation case it takes a slightly different form \cite{aguirre},
namely
\begin{equation}
%G_{\bf q}=\frac{(q_x^2-q_y^2)^2+q^2q_z^2}{q^4}
G_{\bf q}=\frac{q_x^2+q_y^2}{q_x^2+q_y^2+q_z^2}
\end{equation}
Note that this represents a long range interaction with an asymptotic form in real space $\sim 1/r^3$, as the standard Eshelby kernel, yet the spatial symmetry is dipolar in this case.

\begin{figure}
\includegraphics[width=7cm,clip=true]{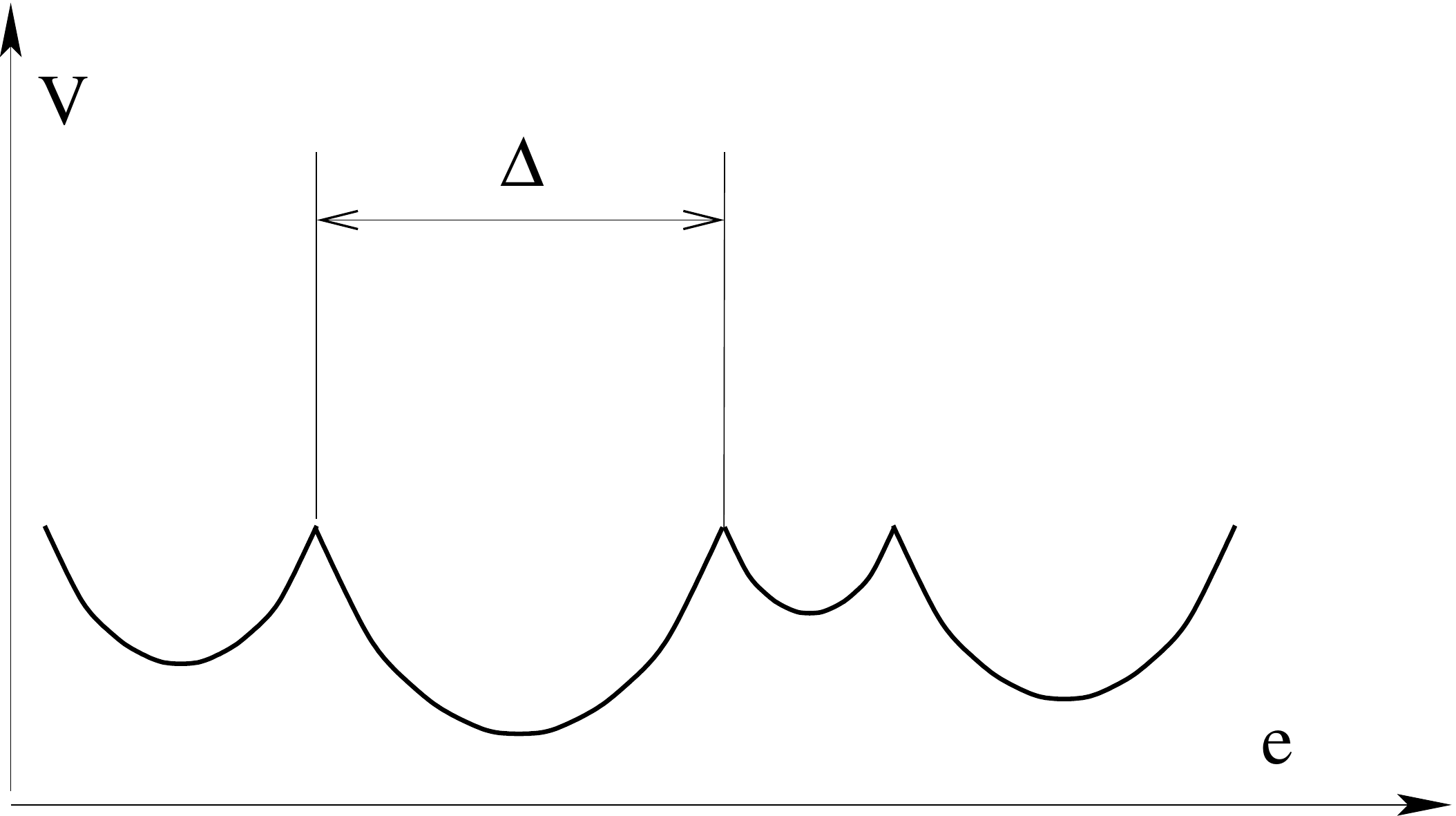}
\caption{Sketch of the local potential for the variable $e$, at some position in the sample. The width $\Delta$ of each potential well is a stochastic variable uniformly distributed between 0.2 and 2.2. Note that different spatial positions in the system have different, uncorrelated forms of this function.
}
\label{sketch}
\end{figure}

The local part of the force $f_{loc}$ models the disordered nature of the system, and it consists of a potential with many minima at different values of $e$, representing different locally stable equilibrium configurations of our amorphous system. The forces $f_{loc}$ are taken totally uncorrelated at different points in the system, namely there are no correlations in the relative positions of energy minima at different points.
The form of the local potentials $V_{loc}(e)$ determining the forces $f_{loc}(e)\equiv -dV_{loc}/de$ is effectively chosen in the following way. At each spatial position we take a local equilibrium value of $e$, namely $e_0$, in such a way that $V_{loc}(e)=(e-e_0)^2/2$, and the force $f_{loc}(e)=(e-e_0)$. When during the dynamical evolution $|e-e_0|$ becomes greater than some value $\Delta$, the local potential well is supposed to destabilize, and the local minimum $e_0$ shifts to a new value. This produces a typical local potential $V_{loc}(e)$ as sketched in Fig. \ref{sketch}. 

The system is driven by externally enforcing the average value of $e$ to be equal to the applied strain $\gamma$.
It has to be stressed that in the present case no additional parameter is included to enforce the driving condition. This is contrary to the standard situation in depinning models where a spring with adjustable stiffness is added to drive the system.
In the simulations shown below $\gamma$ is increased quasi-statically, so each avalanche in the system occurs at a fixed value of strain.

We consider the possibility of relaxation in the system, which is the property that will drive strain localization and "earthquake-like" properties of the avalanches. Relaxation is introduced through a parameter $R$, which represents a back-action of the local forces on the form of the local potentials, in concrete on the values of $e_0({\bf r})$. Relaxation tends to uniformize the stress across the system, and therefore it is accounted for by an evolution of 
the form

\begin{equation}
\dot e_0 =R\nabla^2 (e_0-e)
\end{equation}

The value of $R$ has to be considered in relation to the strain rate $\dot\gamma$. Although we work in the limit of very low $\dot \gamma$, in such a way that strain can be considered constant during the evolution of any single avalanche, there is a strong dependence of the system behavior on the ratio $R/\dot\gamma$, which will be the crucial parameter to describe the dynamics of the model.

\section{Results}

We describe  first the results without relaxation, i.e., the case $R/\dot\gamma=0$.
They are consistent with those obtained with slightly different numerical models (particularly those called elasto-plastic models), although typically in those cases a cubic geometry has been used, instead of the slab geometry we are using here.

\begin{figure}
\includegraphics[width=8cm,clip=true]{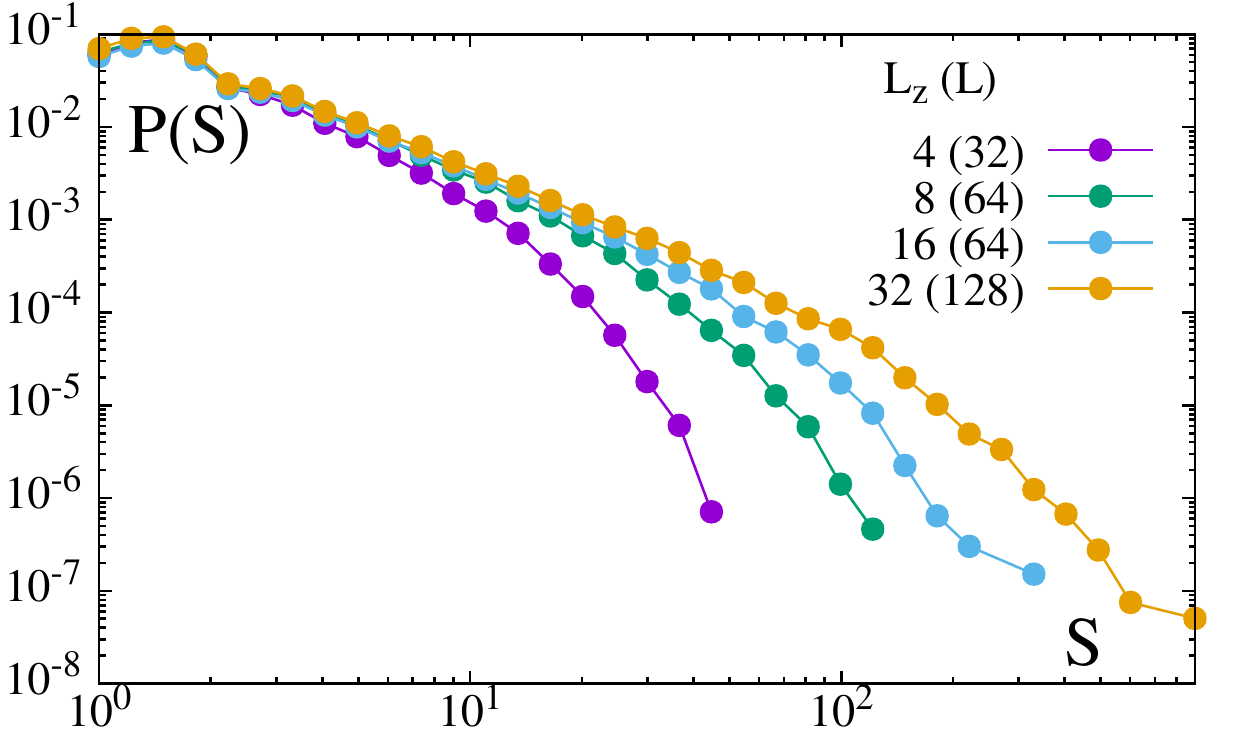}
\includegraphics[width=8cm,clip=true]{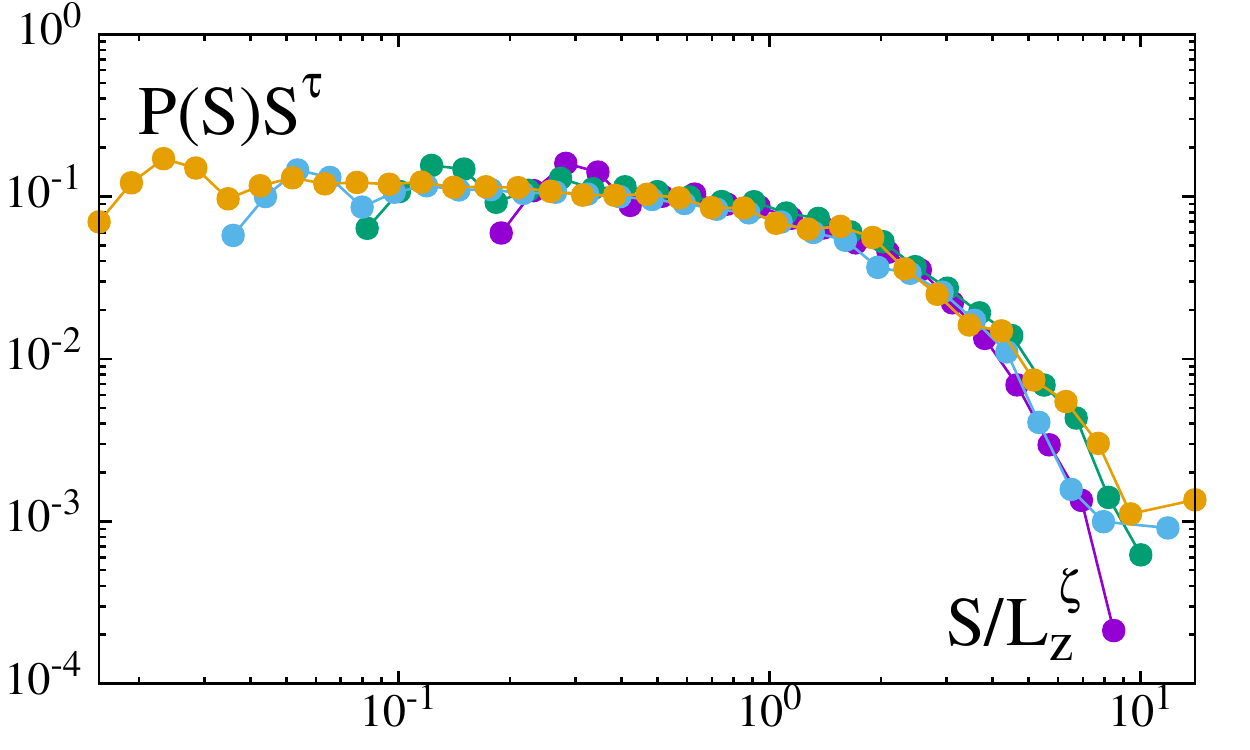}
\caption{(a) Avalanche probability distribution $P(S)$ as a function of avalanche size $S$ for three different values of $L_z$ as indicated (the values of $L$ used along $x$ and $y$ are also reported, but they are irrelevant, because in all cases they are large enough so as to avoid size effects due to $L$). (b) The same data as in (a) but plotted as
$P(S)S^\tau$ vs $S/L^\zeta$, with $\tau=1.5$, $\zeta=1.2$.
% The continuous curve is the function $y=exp(-x/10)$???.
}
\label{avalanchas_r=0}
\end{figure}

In Fig. \ref{avalanchas_r=0}(a) we show the avalanche size distribution as a function of $L_z$. As it was explained before, the value of $L$ along $x$ and $y$ directions is large enough in such a way that these results are not affected by its exact choice. Yet, for completeness, the values of $L$ are also reported in Fig. (\ref{avalanchas_r=0}). As the scaling in panel (b) shows, 
the results can be fitted by a power law 
with a large size cutoff, in the form
\begin{equation}
P(S)\sim S^{-\tau}\exp(-S/S_{max})
\end{equation}
with $\tau\simeq 1.5$, and $S_{max}\sim {L_z}^{\sim 1.2}$. The value $\tau\simeq 1.5$ is the one expected for long range elastic interaction, which in fact is active here due to the structure of the elastic kernel $G$.

\begin{figure}
\includegraphics[width=6cm,clip=true]{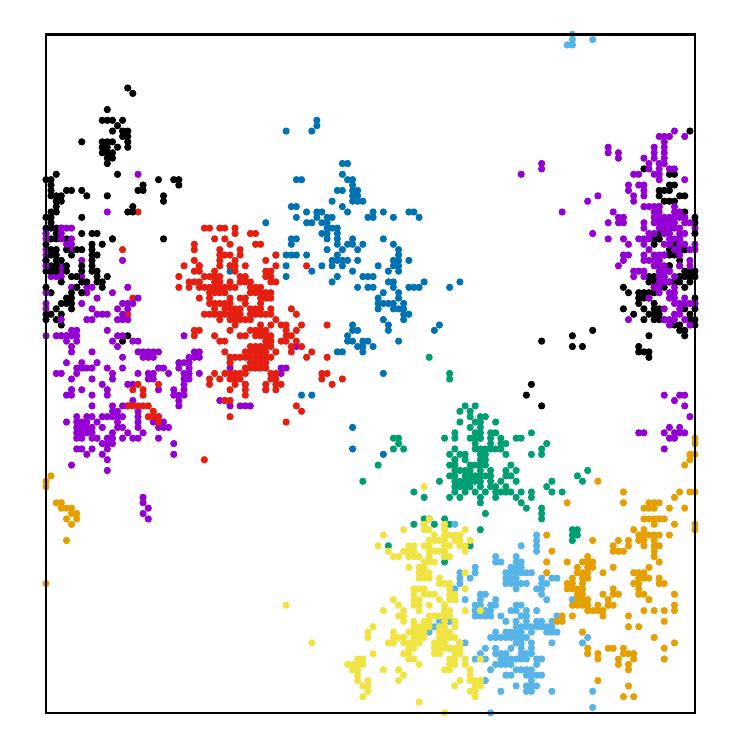}
\includegraphics[width=6cm,clip=true]{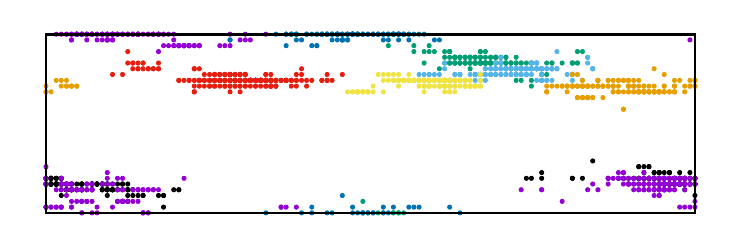}
\caption{Upper and lateral view of the system indicating the sites that were affected by a few successive large avalanches (each avalanche is represented by a different color). Individual avalanches are more or less localized along the $z$ direction, however, successive avalanches appear at random $z$ positions (system size is $128\times 128\times 32$).
}
\label{espacial_r=0}
\end{figure}

The observation of the sites that participate of individual avalanches (Fig. \ref{espacial_r=0}) gives further insight into their structure.
Active sites for individual avalanches tend to be spread along $x$ and $y$ direction, but remain rather localized along $z$. 
The sites participating of an avalanche do not form a connected set,
but typically consist of a number of disconnected pieces. This is an effect that is enhanced by the long range elastic interactions.
When observing a temporal sequence of avalanches, it is seen that they do not occur around the same definite value of $z$, but they rather appear randomly in the system.
Therefore, when considering the deformation induced by many avalanches, this is uniform across the system, with no sign of strain localization. 

Although the largest avalanches in the system diverge in size as $L_z\to\infty$, the $x$-$y$ span becomes actually 
progressively smaller compared with $L_z$ as $L_z$ increases. In fact, we observed that the number of positions in the $x$-$y$ plane affected by an avalanches behaves similarly to the avalanche size itself. \cite{footnote2}
Then 
the span along the $x$-$y$ plane can be estimated as $\sim S_{max}^{1/2}$, which increases sub-linearly with $L_z$ (since $S_{max}\sim L_z^{1.2}$).
This implies that avalanches become ``vanishingly small" compared with system thickness as this is increased, and in the large thickness limit the yielding of the system will be uniform and smooth. We will come back to this fact when comparing with the results for $R/\dot\gamma>0$.

\begin{figure}
\includegraphics[width=6cm,clip=true]{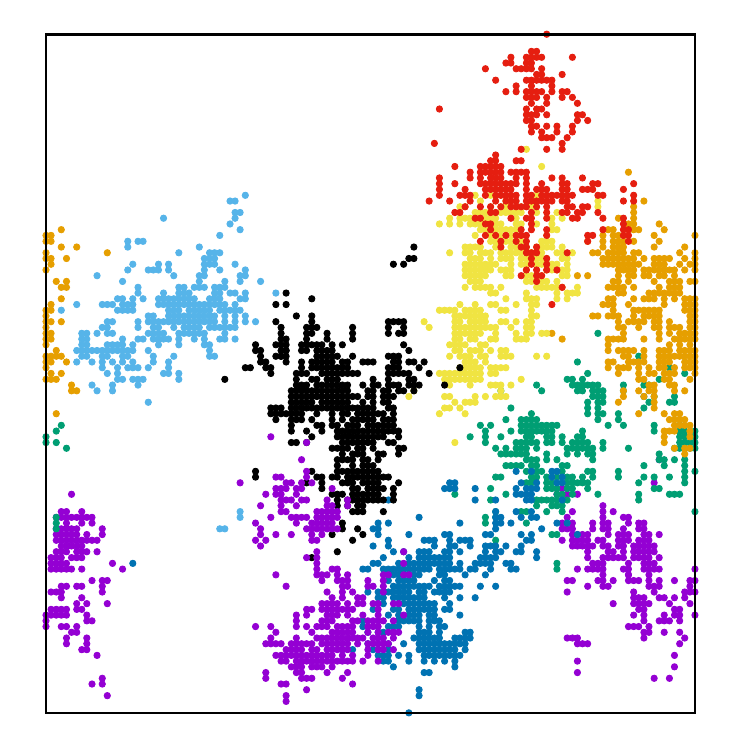}
\includegraphics[width=6cm,clip=true]{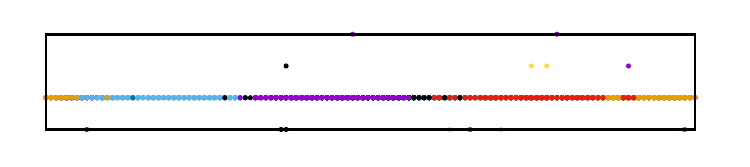}
\caption{Same as Fig. \ref{espacial_r=0} for finite relaxation ($R/\dot\gamma=0.3$). 
The main difference observed is the localization of the deformation around a ``fault" that is formed near a definite $z$ value (system size is $128\times 128\times 16$).
}
\label{espacial_r.gt.0}
\end{figure}

\begin{figure}
\includegraphics[width=6cm,clip=true]{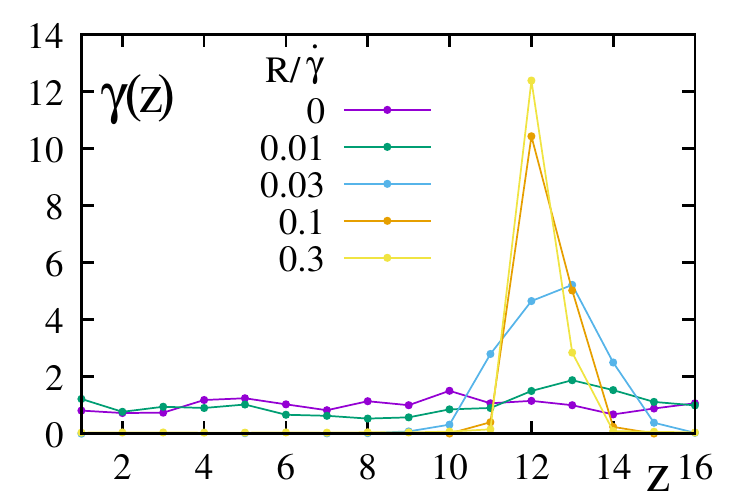}
\caption{Normalized accumulated deformation $\gamma(z)$ at each $z$ value during a long run, as a function of the relaxation parameter $R/\dot\gamma$. The deformation is seen to be uniform when $R=0$, but it becomes localized when $R/\dot\gamma\ne 0$.
}
\label{localizacion}
\end{figure}

Now we present the results obtained by including relaxation in the model. In this case the system is characterized by a competition between the rate of local relaxation and the strain rate. Namely, the control parameter will be $R/\dot\gamma$. 
%(Note that the development of an avalanche is considered as ``instantaneous", namely the strain is kept constant and the relaxation term does not have any effect in the time scale of individual avalanches).
In Fig. \ref{espacial_r.gt.0} we plot the sites affected by a few consecutive avalanches with a size larger than some minimum. 
One remarkable feature that is observed at finite relaxation compared to the unrelaxed case is that deformation is not any more uniform in the system, but localizes in a nearly two dimension ``fault" at some particular value of $z$. 
This is qualitatively observed in the localization of the avalanches as observed in the lateral view in Fig. \ref{espacial_r.gt.0}, and it is also more quantitatively established in Fig. \ref{localizacion}, where we plot the accumulated deformation that has affected each individual layer of the system within a long run. While this deformation is seen to be uniform when $R=0$, it is clearly localized in $z$ when $R/\dot\gamma\ne 0$.

This localization effect is reminiscent of the shear band localization observed in cases of finite $\dot\gamma$. However in this case the argument of a ``Maxwell construction" does not apply, as we do not have uniform sliding within the shear band. Yet the reason to have localization is rather similar: the region of the system in which avalanches occur maintain a state of low relaxation, which produces a local slower critical stress, whereas the rest of the system which is fully blocked had plenty of time to relax and therefore it has a large critical stress. As a consequence, deformation continues to occur systematically at the same spatial positions. In this respect, it has to be mentioned that this localization process is highly hysteretic when the value of $R/\dot\gamma$ is changed. The curves shown in Fig. \ref{localizacion} where obtained starting at $R/\dot\gamma=0$ and progressively increasing it. If, from the final large value of $R/\dot\gamma$ this is reduced back to zero, it is very difficult to return to a situation of uniform deformation, because blocked parts of the sample have already a large critical stress, and are not ready to flow even if relaxation is suppressed. 
The recovery of a uniform deformation occurs via a very slow process that was quantitatively described in \cite{jstat,falk1,falk2} in the context of shear bands at finite strain rate.

\begin{figure}
\includegraphics[width=9cm,clip=true]{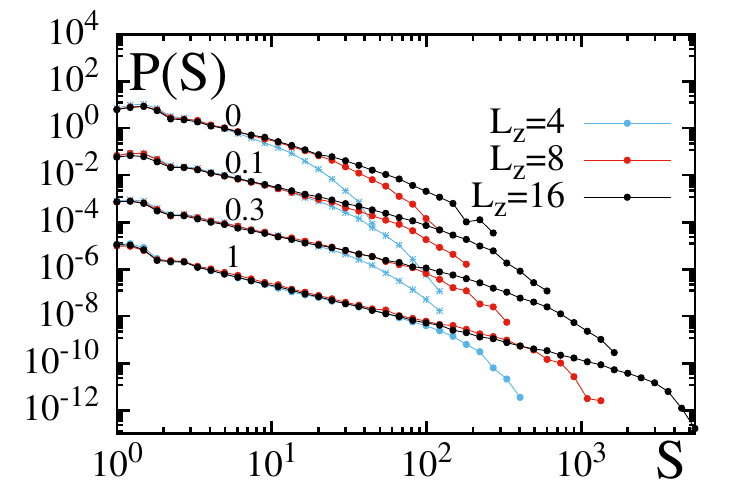}
\caption{Avalanche probability distribution $P(S)$ as a function of avalanche size $S$ for three different values of $L_z$ and different values of $R/\dot\gamma$ as indicated (curves corresponding to different values of $R/\dot\gamma$ were vertically displaced, for better comparison).
}
\label{avalanchas_r}
\end{figure}

\begin{figure}
\includegraphics[width=9cm,clip=true]{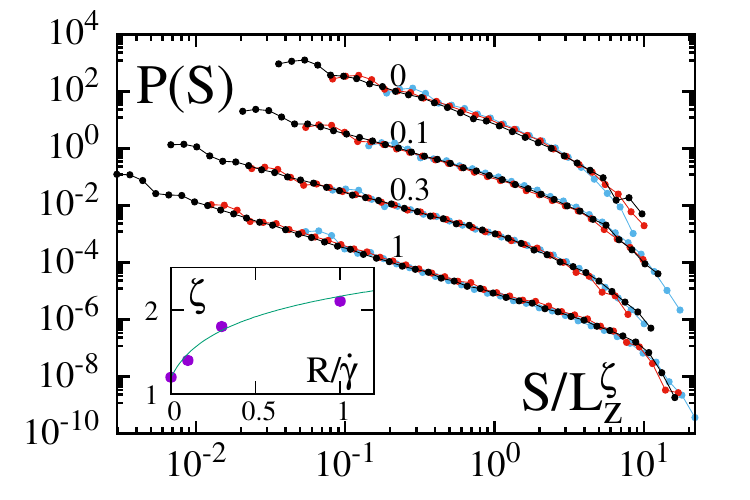}
\caption{The data of the previous figure scaled according to $S/L_z^\zeta$. 
The values of $\zeta$ used to scale data for different values of $R/\dot\gamma$ 
are indicated in the inset. Line is a guide to the eye.
}
\label{escaleo}
\end{figure}

We now analyze how the avalanche size distribution is affected by relaxation. In Fig. \ref{avalanchas_r} we show size distributions for different values of $L_z$ and $R/\dot\gamma$ (here again, $L_x$ and $L_y$ are taken to be large enough, so as not to affect the results).
By comparing curves with the same value of $L_z$, we see that in the presence of relaxation the power law distribution is maintained, roughly with the same value of $\tau$, but the increase of $R/\dot\gamma$ makes the avalanche distribution broader, with a value of $S_{max}$ that increases with $R/\dot\gamma$.
In Fig. \ref{escaleo} the avalanche distributions are plotted as a function of $S/L_z^\zeta$, where the values of $\zeta$ are chosen to collapse the curves corresponding to the same $R/\dot\gamma$, namely, we have an $R$-dependent $\zeta$ value. This dependence is shown in the inset.

We see that $\zeta$ shows a strong increase as a function of $R/\dot\gamma$. It starts at $\zeta\simeq 1.2$ for $R=0$, reaching and even overpassing the value $\zeta=2$ when $R/\dot\gamma \sim 1$. The value $\zeta=2$ is somewhat of a 
critical value, in the following sense. 
Since the linear size of an avalanche in the $x$-$y$ plane can be estimated as $\sim S^{1/2}$, we see that if $\zeta<2$, the maximum linear size of avalanches becomes negligible compared to the system thickness $L_z$ as $L_z$ increases. On the contrary, if $\zeta>2$ this maximum size stays larger than $L_z$ in the thermodynamic limit. We conclude that there is a critical value of $R/\dot\gamma$ below which avalanches become vanishingly small when $L_z\to\infty$, whereas above that value there are avalanches that have a linear extent in the $x$-$y$ plane that is much larger than $L_z$. We suggest this is an indication (in the large system size limit) of a transition between smooth, ductile yielding, and a fragile behavior in which the effect of individual avalanches continues to be observable even if $L_z$ is increased arbitrarily.  This is one of the main findings of the present work.

\section{Roughness of the spatial distribution of stress}

In the previous section we showed that avalanches become vanishingly small in the thermodynamic limit ($L_z\to \infty$)
in the absence, or for low values of relaxation, whereas there are ``macroscopic" avalanches (linear size larger than $L_z$) for large enough relaxation. We will explore the consequence of this fact in the stress distribution across the sample.

\begin{figure}
\includegraphics[width=15cm,clip=true]{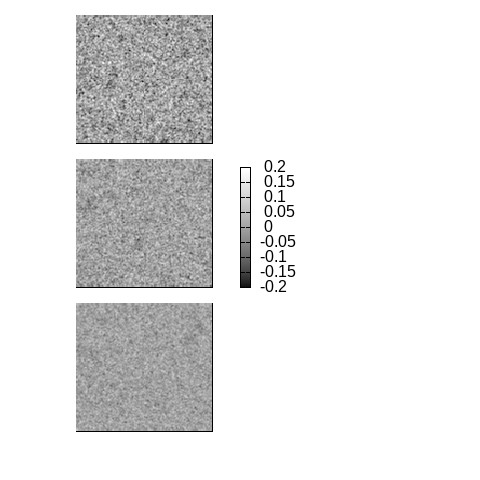}
\caption{Instantaneous stress fluctuation across the sample, for different values of $L_z=8$, 16, 32 ($L_x=L_y=128$), in the absence of relaxation.
}
\label{fotosr0}
\end{figure}

\begin{figure}
\includegraphics[width=7cm,clip=true]{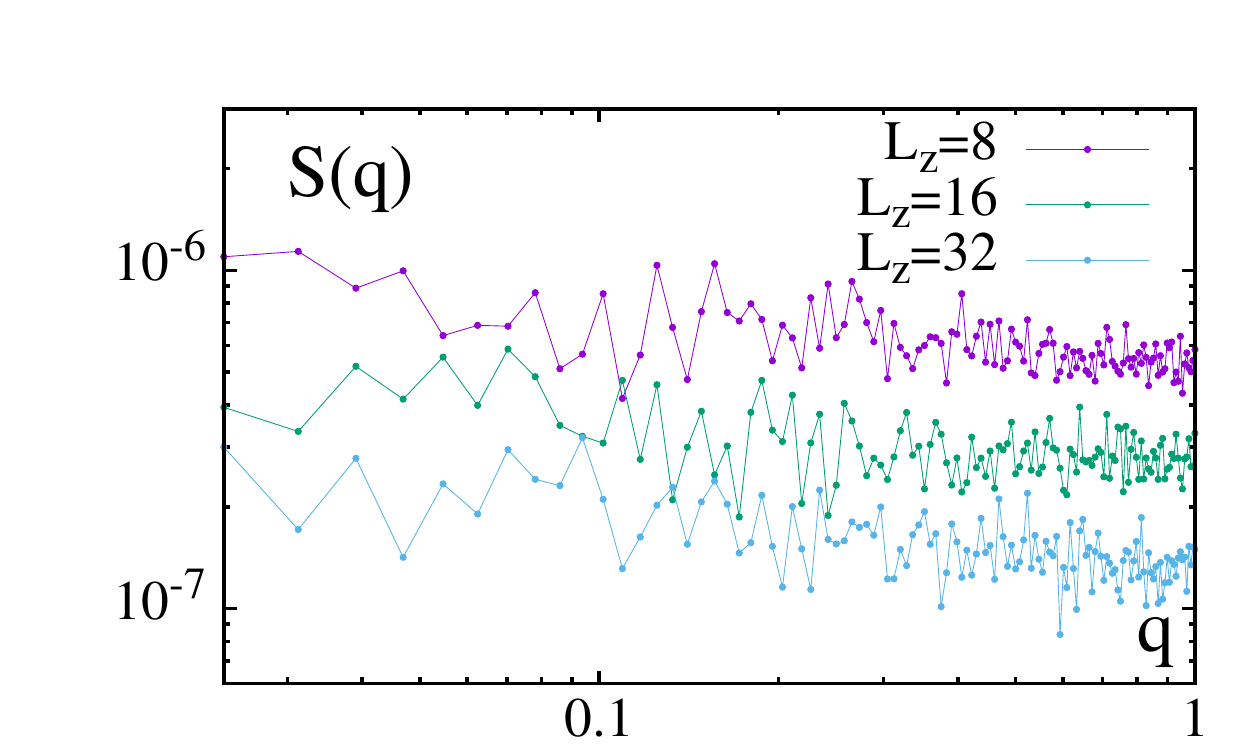}
\caption{Structure factor $S(q)$ of the configurations in Fig. \ref{fotosr0} (``$q=1$" corresponds nominally to the maximum $q$-value allowed by the discrete mesh). The structure factor is flat, and its value decreases as $~L_z^{-1}$.
}
\label{fstruct0}
\end{figure}

In the simulation we have access to the local value of stress at every time, which is given by $\sigma({\bf r},t)=e({\bf r},t)-e_{0}({\bf r},t)$. To be able to extract some manageable information from this, we do the following. First, we work with an average of this quantity over the $z$ coordinate. This generates a function (that we continue to call $\sigma(x,y,t)$ for simplicity)
that is much simpler to analyze, and also eliminates the large differences that would certainly appear between points close or away the yielding plane (at least for cases where spatial localization occurs). 
Therefore we take the function $\sigma(x,y,t)$ at fixed times as an indication of the spatial fluctuation of stress.

As before, we first consider the $R=0$ case. In Fig. \ref{fotosr0} we see the function
$\sigma(x,y)$ at some fixed time for systems with different values of $L_z$. An examination of these figures suggests that fluctuation of  $\sigma$ across the system is reduced as $L_z$ is increased, which is compatible with that stated in the previous section. In order to have a more quantitative confirmation of this behavior, we calculate the structure factor of these configurations, and plot the results in Fig. \ref{fstruct0}. We see a rather flat structure, indicating the lack of correlations of $\sigma$ among different parts of the sample. Moreover, the typical value of the structure factor decreases as $L_z$ increases, indicating that the stress distribution become progressively more uniform in thicker samples. We find a decay of the typical fluctuation of stress $w$ across the sample that follows $w\sim L_z^{-1}$.

\begin{figure}
\includegraphics[width=15cm,clip=true]{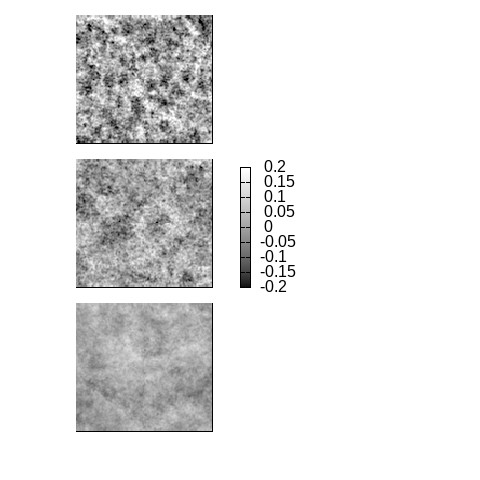}
\caption{Same as Fig. \ref{fstruct0} for $L_z=4$, 8, 16 (we depict a 128$\times$ 128 piece of a system with $L_x=L_y=512$), for $R/\dot\gamma=1$.
}
\label{fotosr1}
\end{figure}

\begin{figure}
\includegraphics[width=7cm,clip=true]{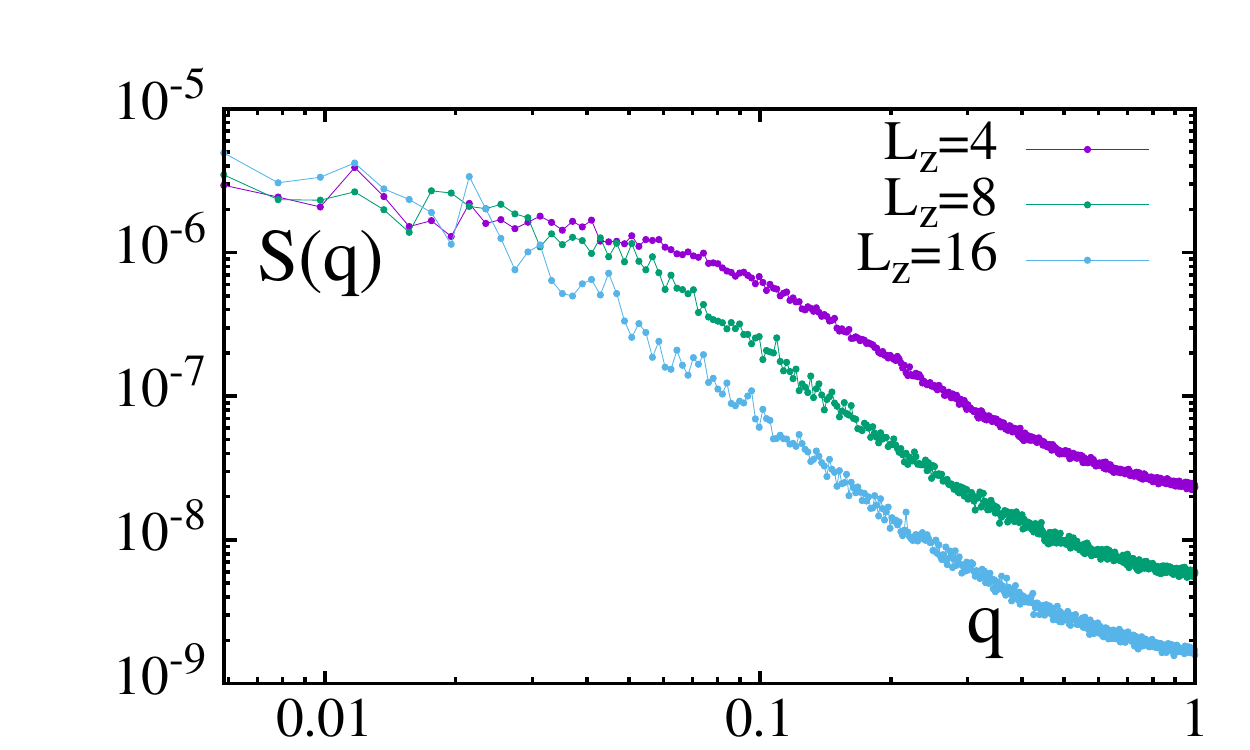} 
\caption{Structure factor of the configurations in Fig. \ref{fotosr1} (an average over many equivalent  configurations was performed). Now the structure factor reaches a $L_z$-independent value as $q\to 0$.
}
\label{fstruct1}
\end{figure}

Now we do the same analysis in a case with $R/\dot\gamma=1$. We see the distribution of $\sigma$ across the system in Fig. \ref{fotosr1}. It is apparent that in the present case there are noticeable correlations between the values of $\sigma$ at different spatial positions. To be more quantitative about this point we resort again to the analysis of the structure factor. The curves corresponding to the three values of $L_z$ (obtained averaging a large number of configuration to minimize the statistical error) are shown in Fig. \ref{fstruct1}.
For large $q$ values, the structure factor decreases with $L_z$, as before. However, for the lowest values of $q$ a remarkable independence of the structure factor on $L_z$ is observed. 
This independence occurs in a range of small $q$ that corresponds to a spatial region of the order of $L_z$, and it is related to the correlation that is visible in the snapshots in Fig. \ref{fotosr1}. We should stress however, that the independence of $S(q)$ with $L_z$ observed at low $q$ does not imply that the stress distribution has a finite width in the limit $L_z\to\infty$.
%{\footnote{For this to be the case, $S(q)$ should increase with $L_z$. Explain}}. 
In fact,
our model does not sustain persisting long range correlations as the relaxation mechanism is a diffusive process that tends to uniformize the stress in the system. In other words, if we go to a limit in which $R/\dot\gamma\to\infty$ (for instance, setting $\dot\gamma=0$  while $R$ remains finite) the system configuration will evolve towards a state of uniform stress. 
Yet the correlation in stress that are observed originate in the existence of large (linear size $\sim L_z$) avalanches in the presence of relaxation. In the next section we correlate this effect with the existence of ``velocity weakening" in the system.

\section{Flow curve and velocity weakening}

The spatial correlations of stress we observe for finite $R/\dot \gamma$ are a consequence of the velocity weakening at the fundamental level caused by relaxation. Yet in the thermodynamic limit  velocity weakening cannot be observed, in the same way that a reentrant liquid-gas isotherm of a system displaying a liquid-gas transition is not really observed, since it is screened by the coexistence of phases. However in finite systems it can be observed, and this is what we want to show here.

\begin{figure}
\includegraphics[width=7cm,clip=true]{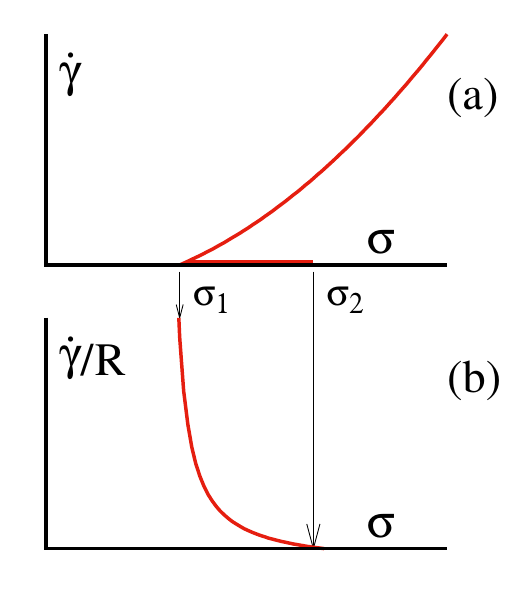}
\caption{(a) Schematic flow curve of a finite size system, displaying a velocity weakening region at very low $\dot\gamma$. In the simulations in the present paper we deal with quasi-static simulations, that correspond to the $\dot\gamma\to 0$ limit. (b) By zooming in the quasistatic limit we have access
to the effect of finite relaxation, characterized by the $R/\dot\gamma$ ratio. This is the region that we explore in the simulations.}
\label{sketch_fc}
\end{figure}

\begin{figure}
\includegraphics[width=7cm,clip=true]{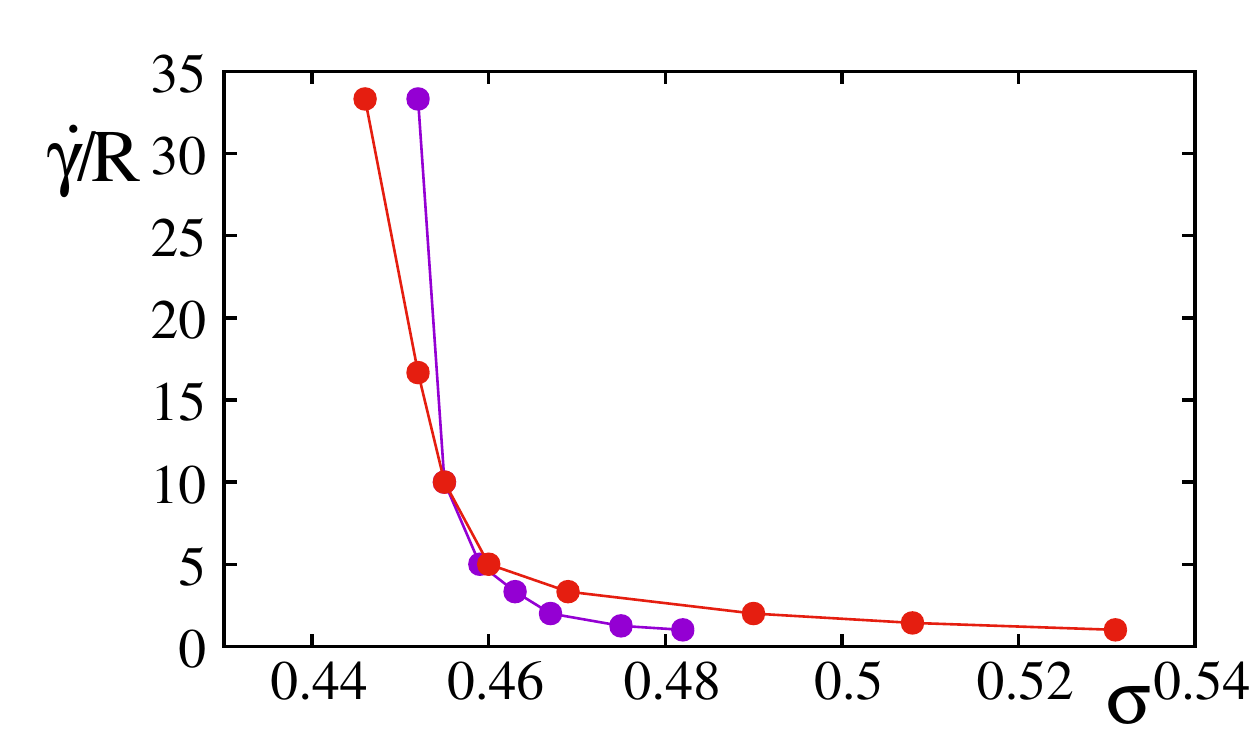}
\caption{Flow curve of our quasi-static simulations showing the velocity weakening effect. Results for $L_z=4$ (red) and $L_z=16$ (blue). Note how the velocity weakening effect becomes less pronounced for the thicker sample.}
\label{flowcurve}
\end{figure}

As it was explained in the introductory section, the present simulations are run in a quasi-static limit in which avalanches are instantaneous events. This means that we are exploring the $\dot\gamma\to 0$ limit of the full flow curve of the system, as indicated in Fig. \ref{sketch_fc}(a). The different scenarios that we discussed as a function of $R/\dot\gamma$ occur within this limit, and can be observed by ``zooming in" the region of $\dot\gamma\to 0$. Doing this, we have access to the region shown in Fig. \ref{sketch_fc}(b). 
Now, although we have considered the $\dot\gamma\to 0$ case, we still have as an independent parameter the ratio $R/\dot\gamma$. To discuss the flow curve it is more instructive to think of $R$ as fixed, and  $\dot\gamma$ as variable. By collecting results from our simulations we constructed the plot in Fig. \ref{flowcurve}.
The velocity weakening effect in this curve is clearly visible. Yet the effect becomes less visible as $L_z$ is increased. This is not surprising.  A negatively sloped velocity-stress dependence is a mechanically unstable situation that at finite strain rates is screened by the breakdown of spatial homogeneity. In fact, the localization of deformation along the $z$ direction that was described in the previous section is a manifestation of this effect, where the unstable behavior is cured through a ``Maxwell construction mechanism" in which most of the system is blocked, and only a very thin region yields. 
But even when the deformation is localized in a single layer (which plays the role of a ``seismic fault") the yielding may have a velocity weakening character that the system will screen by
producing strong inhomogeneities in the $x$-$y$ plane. This is actually the reason of the strong fluctuations we have observed in the previous section for the stress across the system.

This result reinforces the idea of two different yielding regimes in the thermodynamic limit, depending on the extent of relaxation.
One is the smooth regime, occurring for low or no relaxation, in which the effect of individual avalanches is washed out in the thermodynamic limit, and the stress distribution becomes asymptotically uniform. The second regime, occurring at large relaxation, displays some avalanches that are larger in linear size that the system thickness $L_z$, no matter how large this is, and also spatial fluctuations in the values of stress that are intimately related to the velocity weakening nature of the yielding at short scales. 

\section{Summary and Conclusions}

In this work I have presented numerical simulations of a model yield stress material under quasistatic deformation.
The aim was to investigate under which circumstances the deformation is smooth and uniform
at large scales, or jerky and localized. The main result is that the answer to this question depends on the amount of ``relaxation" that is included in the model. This relaxation also affects other characteristics of the model such as the global flow curve.

For no or low levels of relaxation the avalanches that are responsible of the plastic deformation of the system scale with system size in such a way that they become comparatively small as the system size increases. Also, they appear all across the system and the observed deformation is spatially uniform and smooth in the large system size limit. In this case the flow curve of the material is monotonous.

However, when large levels of relaxation are present, the deformation localizes in a very thin layer. In addition, the maximum size of the avalanches observed are comparable to the system thickness no matter how large this is, meaning that individual avalanches may have macroscopic effects even for very large system sizes. All this phenomenology is associated to an underlying velocity weakening behavior of the system, which is the responsible for both the localization of the deformation in a thin ``fault" and the fact that avalanches within this fault have noticeable effects at the system size scale.
 
\section{Acknowledgments}

I thank Alberto Rosso and Giuseppe Petrillo for discussions on a related model that derived in the preparation of this work.


\begin{thebibliography}{1}
\bibitem{fragile}The meaning of fragile here is somewhat different than usual. Usually fragile means that a sample will break suddenly when deformation is increased. In the present case, the sample does not break, but individual avalanches of plastic rearrengments produce observable effects even in samples of arbitrarily large size.
\bibitem{footnote1}Note that we have stated that we work in the limit $\dot\gamma\to 0$. The fact that $R/\dot\gamma$ is finite means that relaxation is also a very slow process in the system. In other words, we can say that during the duration of any avalanche, $R$ and $\dot\gamma$ do not have any effect on the system.
\bibitem{footnote2}This occurs because typically no more than a single site with the same $x$-$y$ coordinates participates in an avalanche, and additionally because every participating site typically moves to the next potential well of the local potential, and no more than that.
\end{thebibliography}
\end{document}